\documentclass[twocolumn,superscriptaddress,prb,showpacs]{revtex4}
\usepackage[dvips]{graphicx}
\usepackage[T1]{fontenc}
\usepackage{amsmath}
\usepackage{amssymb}

\begin{document}

\title{Measurement of refractive index and equation of state in dense
He, H$_2$, H$_2$O, Ne under high pressure in a diamond-anvil cell}

\author{A. Dewaele}
\affiliation{DIF/Département de Physique Théorique et Appliquée, CEA,
BP 12, 91680 Bruy\`eres-le-Ch\^atel, France }
\author{J. Eggert}
\altaffiliation{Present address : Lawrence Livermore National
Laboratory, Livermore, CA 94550} \affiliation{DIF/Département de
Physique Théorique et Appliquée, CEA, BP 12, 91680
Bruy\`eres-le-Ch\^atel, France }
\author{P. Loubeyre}
\affiliation{DIF/Département de Physique Théorique et Appliquée, CEA,
BP 12, 91680 Bruy\`eres-le-Ch\^atel, France }
\author{R. Le Toullec}
\affiliation{DIF/Département de Physique Théorique et Appliquée, CEA,
BP 12, 91680 Bruy\`eres-le-Ch\^atel, France } \affiliation{Universit\'e
Pierre et Marie Curie, Physique des Milieux Condens\'es, T13, E4, B77,
4 Place Jussieu, 75252 Paris Cedex 05, France }
\date{\today}

\begin{abstract}
We present an accurate determination of the refractive index of hydrogen,
 helium, H$_2$O and neon up to 35 GPa at ambient temperature. The experimental method
is based on the combination of two interferometric signals of the
Fabry-Perot cavity containing the sample in the diamond anvil cell.
The data are put in perspective through the variation of the molecular
polarisability with density. Significant electronic changes are
observed. Interesting possibilities of the method are also
illustrated:  the high precision  to finely probe phase transition;
the simultaneous measurement of the volume to obtain the equation of
state of transparent media with around 1\% accuracy.
\end{abstract}

 \maketitle

\section{Introduction}
There exist considerable evidences that pressure can dramatically change
the electronic properties of a system. On the atomic scale, the
variation of the electronic kinetic and potential energies with volume
tends to make a contraction of the atomic size and a delocalisation of
valence electrons. Consequently, spectacular transformations have been
observed, such as rehybridization of atomic orbitals, like in carbon
(sp$^2$) to (sp$^3$) across the graphite to diamond transition ;
insulator-metal phase transitions, like in O$_2$~\cite{Desgreniers90},
Xe~\cite{Eggert89} or CsI~\cite{Eremets98}; the transformation of ice
from an hydrogen bonded solid to an ionic solid in symmetric
ice~\cite{Goncharov96}. Most changes of the properties of a
system under pressure are related to the microscopic electronic
reorganization, and the determination of the refractive index over a large
frequency domain is probably the most direct method to probe the
electronic structure. The refractive index expresses the response of
the distribution of charges to a perturbing electric field. In order to
extract a detailed change of the charge distribution of a system with
pressure through the measurement of the refractive index, a large
frequency domain has to be investigated. This is unfortunately very
difficult at high pressure. However, already interesting electronic
evolutions have been obtained from measurements of the refractive index
performed in the visible, for instance in hydrogen with the red shift of an
electronic transition that parallel the closure of the electronic gap
(see references below).

Also, an accurate determination of the refractive index under pressure has
a clear metrological importance. In Brillouin spectroscopy,
where the refractive index of the sample and of the pressure
transmitting medium have to be known to extract the sound
velocity from the frequency shift~\cite{Grimsditch88}. In the newly developed laser shock measurements
on pre-compressed targets to extract the shock front velocity
from the phase shift of the VISAR (velocity interferometric
system for any reflector) diagnostics~\cite{Lee02}.
Finally, to measure easily the
thickness of the sample chamber in the diamond-anvil cell
(DAC) with a single spectrometer, when the sample or the
pressure transmitting medium is transparent. This can be
directly used to measure the Equation of State (EoS) of the
sample (see below), or help to control the experimental
conditions, for instance by evaluating the strain of the
studied sample.

In the DAC, the tips of the diamond anvils that enclose the sample form
a Fabry-Perot interferometer. This geometry have already been exploited
to measure the refractive index of transparent samples at high
pressure. Most of the experimental methods developed so far have been
based on the measurement of fringes contrast (\cite{VanStraaten82},
~\cite{VanStraaten88}, ~\cite{Hemley90b}, ~\cite{Evans98}). Other were
based on a controlled sample thickness
(\cite{Eggert92},~\cite{Balzaretti93}). In one work~\cite{LeToullec89},
the combination of the measurement of parallel white beam interference
spectrum and convergent monochromatic beam inferference rings allowed
the determination of both refractive index $n$ and thickness $t$ of the
sample. We present here an improvement of this method.

      With this setup, we have also been able to measure the Equation
of State of helium, by estimating the sample chamber volume changes
upon compression. Such a method has been previously described for H$_2$
~\cite{VanStraaten88}, ~\cite{Evans98}. We show here that with
numerical photograph technology and the accurate determination of the
refractive index, its accuracy becomes comparable to other techniques.
Therefore, it becomes an interesting alternative to X-Ray methods
(X-Ray diffraction and X-Ray absorption~\cite{Shen02}) for lights
elements, particularly light liquids, or to adiabatic compressibility
measurement of Brillouin spectroscopy~\cite{Grimsditch96}.

Three goals have mainly motivated the present accurate
refractive index measurements on four systems, He, H$_2$, Ne
and H$_2$O. First,  H, He and H2O are the main constituents
of planetary interiors and the determination of their
properties up to planetary core conditions is motivating a
lot of effort.  Second, He and Ne   are the most commonly
used quasi-hydrostatic pressure transmitting media in DAC. In both cases, the determination of the refractive
index should be useful to analyze some spectroscopic
measurements. Finally, they have simple electronic structures
with different electronic organisations and the present data
set could be interesting  to test the confidence of {\it
ab-initio} calculations in predicting the optical properties.

The experimental method and a discussion of the uncertainties is
presented in section II. The refractive index measurements are
presented in section III. The analysis of the data in terms of the
changes in the polarizability with density is presented in section IV.
Applications of the method to the determination of the equation of
state and to the study of phase transitions are described in section V.

\section{Experimental procedure }

\subsection{High pressure techniques}
In these experiments we used a membrane diamond anvil cell described
elsewhere (\cite{Letoullec88}) with a large optical aperture ($\pm$
30$^\circ$) and equipped with 400 or 500 $\mu$m table diamonds. Samples
have been loaded in a rhenium gasket at room temperature, by the use of
high pressure loading techniques. In the case of H$_2$O, ultra-pure
water was inserted in the gasket hole using a syringe. As explained
below, thick samples are needed for a better accuracy of the refractive
index measurements. For this reason, double rhenium gaskets have been
used in the highest pressure runs. These gaskets are
 prepared in the following manner: two rhenium gaskets are
pre-indented, then polished on one side and glued together.
With this method, the thickness of an helium sample was of
the order of 50 $\mu$m at 200 kbar, 30 \% more than with a
simple gasket. The pressure gauge was ruby, calibrated with
the quasi-hydrostatic pressure scale~\cite{Mao86}, that
allows pressure measurement with an accuracy of the order of 0.5 kbar.
The absolute accuracy is the one of the ruby scale, {\it i.e}
 at maximum 2\% over the pressure range investigated here.

\subsection{White light interferences}
\begin{figure}[tp]
\includegraphics[height=.3\linewidth,clip=]{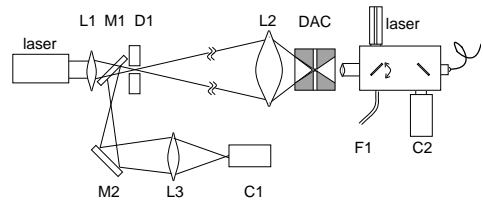}   
\caption{Experimental setup. The light emitted by the laser 1 (He-Ne, 7
mW or Ar) is extended by L1 (simple lens, $f=30$ mm) and focused on the
sample by the objective L2 ($f=50$ mm) ; interference rings propagate
through L2, M1, M2, L3 (objective, $f=80$ mm) and are recorded by the
digital camera C1. On the right, the PRL setup (doubled YAG laser, F1:
illuminating optical fiber, C2: digital camera, S: spectrometer optical
fiber) equipped with a $ \times 20$ Mitutoyo objective, allows pressure
measurement, imaging of the sample and white light interferences
measurement. \label{fig:Exp}}
\end{figure}
The refractive index setup is modified from LeToullec {\it{et
al.}}'s one (see Fig.~\ref{fig:Exp}). The white light
interference pattern in transmission is recorded over the
range 500nm to 700nm with a double substractive spectrometer
(Dilor).
A small area, varying between 20 x 20  and 10 x 10 $\mu$m$^2$, was
selected to perform the measurement. Under high pressures ($P> 250$
kbar), the contrast drastically decreased out of a central flat region,
because of the strain of the diamonds under pressure. The wavelength
difference between two intensity maxima :
\begin{equation}\label{eq:lumblanche}
\Delta \lambda = \frac{\lambda^2}{2 n(\lambda) t},
\end{equation}
where $t$ is the thickness of the sample, $n(\lambda)$ its refractive
index, was determined. In that wavelength range, dispersion could be
observed only in H$_2$O samples; in that case, a linear dependency of
$1/n$ on $\lambda$ was assumed in order to fit Eq.~\ref{eq:lumblanche}.

\subsection{Monochromatic interferences}

The setup presented in Fig.~\ref{fig:Exp} allows the measurement of the
reflective Fabry-Pérot rings produced by the cavity between the
diamonds, when illuminated by a convergent monochromatic beam. We used
an Argon laser ($\lambda =$ 488.0 nm) for helium samples 1 to 3, and a
He-Ne laser ($\lambda = $ 632.8 nm) for H$_2$O, H$_2$, Ne and He
samples.

The interference order $k$ at angle $i$ is defined by:
\begin{equation}\label{InterfOrder}
k = \frac{1}{\lambda} 2 t (n^2-\sin^2i)^{1/2};
\end{equation}
if $k$ is an integer, a minimum in intensity is observed.

At the center of the pattern, the interference order
 $
 k_0={2 n(\lambda) t}/{\lambda},
$ can be deduced from white light interferences measurement. The
angular position $i_p$ of the $p^{{th}}$ intensity minimum has an
interference order $k_p=[k_0]-p+1$. Refractive index and thickness
can be deduced from two fringes ($p$ and $p'$) angular positions:
\begin{equation}\label{index}
n^2=\frac{k_{p'}^2 \sin^2 i_p - k_{p}^2 \sin^2
i_{p'}}{k_{p'}^2-k_p^2} \end{equation}
 and
\begin{equation}\label{thickness}
t^2=\frac{\lambda^2}{4}\frac{k_{p'}^2-k_p^2}{\sin^2 i_p - \sin^2
i_{p'}}
\end{equation}
Interference rings were projected and recorded by
 a CCD camera (C1, see Fig.~\ref{fig:Exp}). The radius $R_p$ of the
rings on C1 is thus measured. Because of optical aberrations, mainly
caused by the objective lenses L2, the relationship between $R_p$ and
$i_p$ was non trivial. To correct these abberations, a reference
spectrum was recorded. This spectrum consisted in fringes formed by an
empty indented and drilled gasket, placed in the DAC. The adjustment of
the optical setup was performed as it is described herebelow. First,
the cell was centered and perpendicularly oriented on the laser beam.
Then M1, C1, L3, L2, L1 and D1 were positioned and the calibration
spectrum recorded. All theses elements were mounted on ($x$, $y$,
$z$, $\theta$) linear and rotation stages. After calibration, all the elements,
except the cell, remained fixed. In the first experiments, the cell was
regularly removed to record the white light interferences pattern. The
best position of the cell (in the $x$ direction) was determined, within
$\pm$ 10 $\mu$m, by optimization of the rings pattern stability with
respect to Y and X translations. This allowed a reproducibility of the
index measurement of $\Delta n = \pm 5.10^{-4}$. Each digital
Fabry-Perot pattern was automatically processed by a program that
evaluates thickness and index for all available pair of rings. This
procedure allows the analysis of numerous spectra, and again,
optimization of the $x$ position of the sample. Tentative error bars
have been estimated for each measurement, taking into account two
factors: first, the uncertainty on rings radius $R_p$; second, the
scatter of the refractive index given by different couples of rings
(Eq.~\ref{index}).

The precision of refractive index measurements is limited by the
following effects:

- the precision of $k_0$ measurement by white light interferences. The
relative precision on $k_0$ estimate is better than 10$^{-3}$ ($\Delta
k < 0.2$) for weakly dispersive media (He, Ne and H$_2$) and of
approximately 2.10$^{-3}$ for dispersive media (H$_2$O). This
measurement must also be carried out in the same sample region than
the region illuminated by the laser beam for monochromatic
interferences, in order to prevent any error caused by non-parallelism
of the diamonds.

- the number of Fabry-Pérot rings. They must be as numerous as possible to statistically
increase the precision of the measurement. In these experiments, a
minimum of four rings has been observed. For this purpose, the angular
aperture of the optical setup was increased to 25$^\circ$. The number
of rings is also proportional to the sample thickness; therefore, thick
samples have been used.

- the elastic strain (cupping) of the diamonds. This is the main source of
uncertainty of our measurements, that limits the pressure range of this
method. This cupping can be first observed by translating the cell in
$y$ or $z$ directions and looking at the modification of the rings
pattern. It can be noted that these measurements constitute an accurate
mean of studying the strain of the anvils. At the beginning of the
compression, inverse cupping, caused by the gasket strength, is
observed. Upon pressure increase, the diamonds become flat and then
cupping occurs. It has been observed approximately at the same
thickness of a simple rhenium gasket (40 to 45 $\mu$m for H$_2$O
samples), and for pressures varying between 100 and 200 kbar. When
pressure increases, it perturbs the rings pattern out of a central flat
zone, up to a point that makes measurements impossible. To reduce this
intrinsic limitation, thick gaskets can be used because they increase
the pressure at which cupping appears. This elastic strain also makes
the focusing diagnostic less precise.

- the pressure gradients in the solid phase. The ruby chip can not be directly located
where the sample was illuminated, because it perturbs the Fabry-Pérot
interferometric rings. It was located at $\simeq$ 60 $\mu$m from the
refractive index measurement point. In separate runs we have estimated
the radial pressure gradient in ice VII, in the same sample geometry;
this gradient is negligible up to 200 kbar and reaches 0.33 kbar/$\mu$m
at 350 kbar. In ice, the pressure was corrected by this factor.
Actually, this pressure gradient is also an intrinsic limitation of
the method, because the extend of the source of interference rings is
100 $\mu$m, if $i<25^\circ$; at high pressure, rings with $i<15^\circ$
are analyzed in order to reduce this error.

\subsection{EoS measurement}
An analysis of numerical images of the sample taken in transmitting
light allows the measurement of the surface $S$ of the sample, with a
precision of 1 \%. The volume $V_{S}$ is then obtained by multiplying
$S$ with the thickness of the sample, obtained at the same time as its
refractive index. The uncertainty on volume determination is mainly
caused by the uncertainty on $S$. No corrections taking into account
the cupping of the diamond was necessary for the measurements presented
in Sec. V, but these corrections are in principle possible. We made the
assumption that the amount of matter trapped in the sample chamber
remained constant upon pressure increase. In that case, the EoS $V(P)$
can be simply deduced from $V_{S} (P)$ by choosing a reference point
$V_0(P_0)$. For a given run, the molar volume of the sample can be
expressed as:
$$
V_{\mathrm{mol}}(P)=V(P)\frac{V_{\mathrm{mol}}(P_{\mathrm{ref}})}{V(P_{\mathrm{ref}})},
$$
$V(P)$ being the volume of the sample at the pressure $P$ and
$V(P_{\mathrm{ref}})$ the volume of the sample at the reference
pressure
 $P_{\mathrm{ref}}$. We noticed that the precision of the volume
measurement increased when the thickness of the sample decreases. A
similar observation has already been made by Evans and Silvera, 1998,
~\cite{Evans98}. This is probably because the geometrical imperfections
of the gasket hole, extensively described~\cite{VanStraaten88},
are smaller for thin samples.

\section{Refractive index data}

\subsection{Helium}
\begin{figure}[tp]
\includegraphics[height=90mm,clip=]{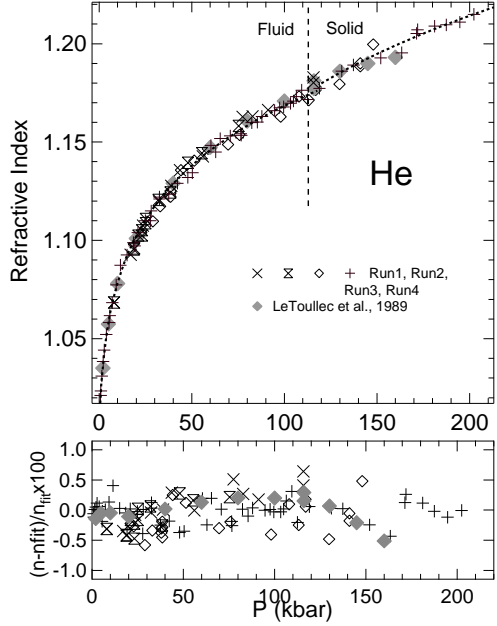}   
\caption{{\bf Upper:} Refractive index of fluid and solid He as a
function of pressure at ambient temperature. The first 3 runs have been
performed at $\lambda = 488$ nm, and the last one at $\lambda = 632.8$
nm. The parameters of the fitted functions (dotted lines) are presented
in Table~\ref{IndexTable}. {\bf Lower:} Difference, in \%, between
experimental data and the fitted function.\label{fig:HeIndex}}
\end{figure}
The refractive index data obtained in 4 runs up to 210 kbar
are shown in Fig.~\ref{fig:HeIndex}. During the same
experimental run, the orders of interference of the rings of the
fringe pattern have been followed between each pressure steps.
The good agreement between the data of the different runs
data indicates that the uncertainty associated to the white
light determination of the order of interference is
negligible. For all data shown in Fig.~\ref{fig:HeIndex}, the
error bars are smaller than the size of the
symbols. 
The difference between the fitted index law and LeToullec measurements
is smaller than 2.10$^{-3}$, except for the highest pressure points
where LeToullec measurements were less precise.

\subsection{Neon}
\begin{figure}[tp]
\includegraphics[height=1.05\linewidth,clip=]{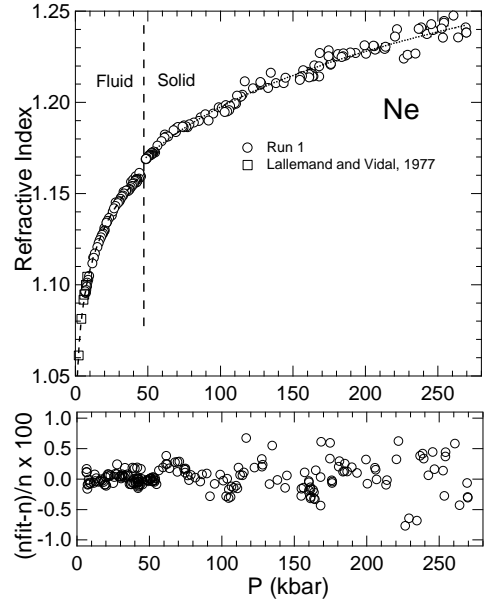}  
\caption{Same plot as Fig.~\ref{fig:HeIndex}, for neon. Experiments
have been performed at $\lambda = 633$ nm. \label{fig:NeIndex}}
\end{figure}
One run has been performed with a neon sample, up to 27 GPa. The
evolution of the refractive index of neon with pressure is plotted in
Fig.~\ref{fig:NeIndex}, The refractive index of Ne had been determined
previously to a maximum pressure of 1 GPa from capacitance measurements
~\cite{Lallemand77}. The present data are in very good agreement with
this previous determination, in the pressure range of overlap.

\subsection{Hydrogen}
\begin{figure}[tp]
\includegraphics[height=1.2\linewidth,clip=]{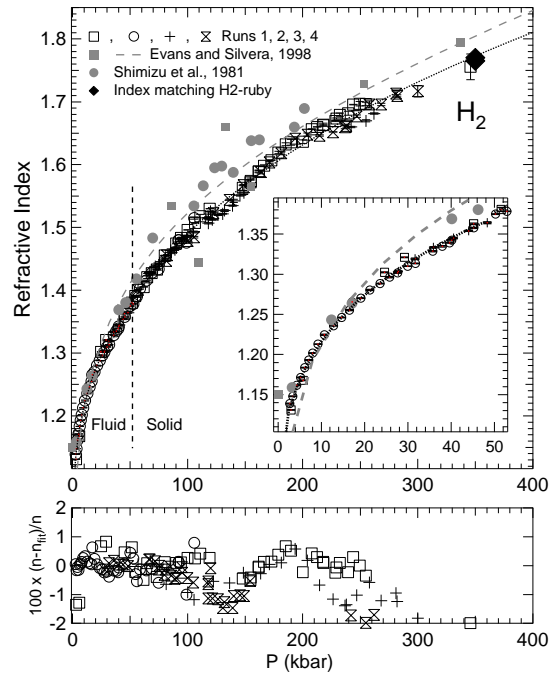}   
\caption{Refractive index of H$_2$ at $\lambda = 632.8$ nm. {\bf
Inset:} enlarged region in the fluid phase. \label{fig:H2Index}}
\end{figure}
 The evolution of the refractive index of hydrogen with pressure is
plotted in Fig. \ref{fig:H2Index}.

There is a systematic difference between our data and
previous determinations, either the one by brillouin
scattering ~\cite{Shimizu81} (1.5 \%) or the one by white
light interference ~\cite{Evans98} (around 1 \%, which is
within their estimated error bars). The matching of
refractive index of ruby and hydrogen has been observed
(annulation of the contrast) at 35 $\pm 10$ GPa. This
provides an anchorage point to the refractive index curve. At
this pressure, the refractive index of ruby is expected to be
1.77 ~\cite{Balzaretti93}; this value is in good agreement
with the present measurements and that gives a good confidence in our
highest pressure runs.
\begin{table}[htb]
\begin{tabular}{ccc}
\hline
Compound  & Refractive index & P range \\
& & (kbar)  \\
\hline
Fluid He &$n = 0.8034 + 0.20256(1+\, P)^{0.12763}$ & 0.8-115 \\
Solid He &$n = -0.1033 +(1+\, P)^{0.052}$ &117-202 \\
Fluid Ne&$ n = 0.67 + 0.33(1+4.3 \times P)^{0.076}$ & 7-47 \\
Solid Ne& $ n = 0.9860 + 0.08578 \; P^{,0.1953}$  &  50-270\\
 Fluid H$_2$ & $ n = 0.949 + 0.06829(1+11.8 \, P)^{0.2853}$&3-49  \\
Solid H$_2$ & $ n = 0.95 + 0.1144 \; P^{0.3368}$& 52-350 \\
Fluid  H$_2$O & $ n = 0.900 + 0.4323(1+0.1769 \, P)^{0.164}$& 0.5-12 \\
Ice VI& $ n = 1.425 + 0.00255 \, P$ & 12.7-22 \\
Ice VII &$ n = 1.175 +0.2615(1+0.101 \, P)^{0.222}$& 30-354 \\
\hline
\end{tabular}
\caption{Fitted forms of refractive index vs. $P$ (in
kbar).\label{IndexTable} }
\end{table}

\subsection{Water and ice}
\begin{figure}[tp]
\includegraphics[height=0.85\linewidth,clip=]{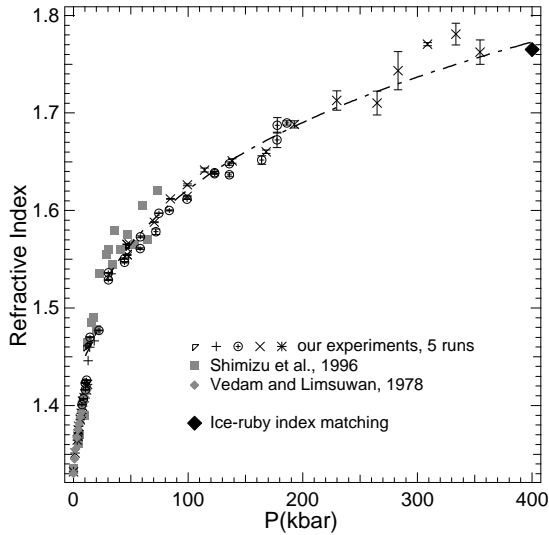}   
\caption{Measured refractive index of water, ice VI and ice VII, at
$\lambda = 632.8$ nm, with the corresponding fitted functions (dotted
lines). \label{fig:H2OIndex}}
\end{figure}
For H$_2$O, 5 experimental runs have been made, up to 35 GPa.
Due to the dispersion of the refractive index of H$_2$O, of
the order of 1 \% over the 500 - 800 nm range, the
determination of the order of interference from white light
measurements is less precise and consequently small
deviations between different experimental runs can be
detected in the fluid phase. However, our measurements agree
very well with precise interferometric measurements in the
fluid phase ~\cite{Vedam78}.  At higher pressures, our data
are within the scatter of a previous Brillouin scattering
determination ~\cite{Shimizu96}. At pressures higher than 20
GPa, the strength of ice causes pressure gradients in the
sample, that can perturb the pressure estimation. To correct
it, the pressure measured by the ruby chip $P_R$ was then
corrected by a factor $\Delta P = d \times \mathrm d
P/\mathrm d r$, $d$ being the distance between the ruby chip
and the center of the sample, and the last factor has been
estimated in a separate run by mapping the sample chamber
with ruby (in kbar/$\mu$m): ${\mathrm d P}/{\mathrm d
r}=(P_{R} \times 0.425)/{t}.$ The estimated error bars
plotted in Fig.~\ref{fig:H2OIndex} increase drastically as
pressure increases, showing the decrease of the quality of
the fringe pattern associated to the strain in the diamonds.
Refractive index matching between ice VII and ruby has been
observed at $P = 40.5 \pm 10 $ GPa. At this pressure, the
refractive index of ruby is approximately 1.765
~\cite{Balzaretti93}, 1.2 \% lower than the extrapolation of
our measurements. That is within the uncertainty of our
highest pressure measurements.

The functions obtained by a fitting procedure to our experimental data
are gathered in table~\ref{IndexTable}. The use of power laws is
justified, to a certain extent, by the fact that the Lorentz-Lorenz
factor defined in Eq.~\ref{LorLor} is approximately constant (see
below). We used as least parameters as possible to fit the data. The
quality of the various fits is shown in the various figures that
present the refractive index of He, H$_2$, Ne, H$_2$O versus pressure.


\section{Discussion}

\subsection{Polarizability}
\begin{figure}[tp]
\includegraphics[height=0.8\linewidth,clip=]{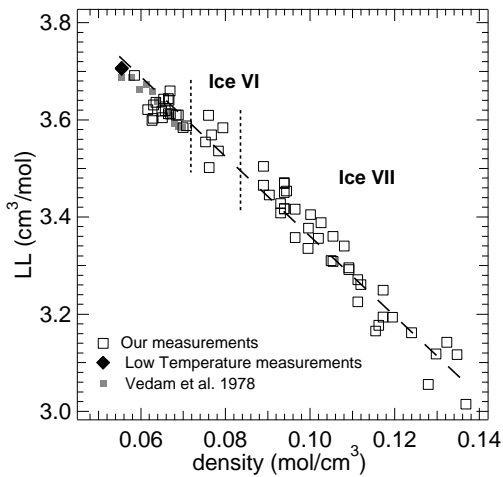}   
\caption{Lorentz-Lorenz factor of H$_2$O phases as a function of molar
density. \label{fig:LLH2O}}
\end{figure}
The refractive index is generally related to the
polarizability of the molecular entities of the system
through the  the Lorentz-Lorenz formulation:
\begin{equation}\label{LorLor}
  LL = \frac{1}{\rho}\frac{\epsilon-1}{\epsilon+2}= \frac{4 \pi N_A}{3}
\alpha.
\end{equation}
 In Eq.~\ref{LorLor}, $N_A$ is the Avogadro number,
$\epsilon$ the relative dielectric constant, and $\alpha$ the molecular
polarizability of the compound. In the static case, this equation is
called the Clausius-Mossoti relation. It arises from the expression of
the local electrical field for dense materials: ${\bf E_{loc}(r)}={\bf
E_{macro}(r)}+4 \pi {\bf P(r)}/3+{\bf E'(r)}$. ${\bf P(r)}$ is the
polarization and ${\bf E'(r)}$ is the field created by the close
neighbors of the electronic cloud centered on ${\bf r}$.
Eq.~\ref{LorLor}, corresponds to ${\bf E'(r)}=0$. This is the case on
symmetry center of a a cubic crystal. For hydrogen, helium and neon,
Eq.~\ref{LorLor} is commonly used~\cite {Evans98},~\cite{LeToullec89}
because they are constituted by spherical molecular entities (in the
case of hydrogen the molecules are freely rotating), organized in
highly symmetric structures. Also, for these systems in the pressure
range covered here, the electronic transitions are far from the
visible. The imaginary part of the refractive index should then be very
small and $\epsilon = n^2$ is a very good approximation.

 In the case of water, the polarizability is generally
written as the sum of three components: the polarizability due to the
orientation of the permanent dipole of the water molecule, the
polarizability due to the electronic changes associated to the inter
and intra- molecular modes and the polarizability associated to the
electronic transitions within the water molecules. The contribution of
the permanent dipoles operates at frequencies below 10$^{5}$ Hz. The
contribution due to the intermolecular modes is important below
10$^{11}$ Hz in the liquid and 10$^{5}$ Hz in the solid
~\cite{Petrenko99}. The contributions due to the intra-molecular modes
should be important below 10$^{14}$ Hz (the highest frequency vibration
mode of H$_2$O is a stretching mode at 3656 cm$^{-1}$ in the gas
phase). Above 10$^{16}$ Hz electronic transitions should be considered.
The optical window covered here is just in between these dispersion
phenomena, which implies a small dispersion of the refractive index and
a small absorption. We did not observe any pressure effect on the
dispersion. In Fig.~\ref{fig:LLH2O}, the Lorentz-Lorenz factor of water
and ice calculated using the present data and published equations of
state of liquid water, Ice VI and Ice VII (\cite{Saul89},
~\cite{Bridgman42}, ~\cite{Hemley87}, ~\cite{Loubeyre99}) is plotted,
as a function of molar density $\rho$. It is interesting to note that
the variation of $LL$ plotted on Fig.~\ref{fig:LLH2O} does not depend
on the microscopic arrangement of the molecules but only of the
density. This tends to prove that the hypothesis ${\bf E'(r)}=0$ in the
local field is also reasonably valid in water. Consequently,
Eq.~\ref{LorLor} should be valid here to analyze the data. It is seen
in figure 6 that the $LL$ factor decreases by approximately 20 \% up to
the maximum pressure of investigation. The approximation $LL$ constant,
that is used to analyse spectroscopic data of water under pressure,
such as the Brillouin scattering sound velocity measurements
~\cite{Grimsditch96}, is not correct because it leads to an
underestimation of the refractive index of the order of 3 \% at 35 GPa.
\begin{figure}[tp]
\includegraphics[height=0.85\linewidth,clip=]{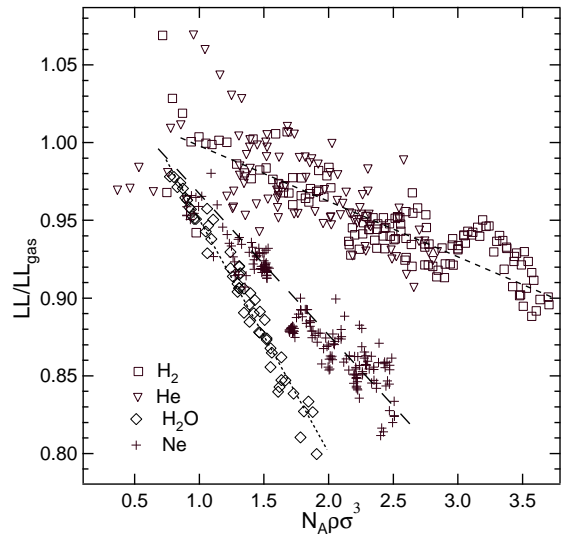}   
\caption{Lorentz-Lorenz functions of the studied compounds, normalized
to their gas
value~\protect\cite{Achtermann93},~\protect\cite{Lallemand77},~\protect\cite{Schiebener90}.
\label{fig:LL}}
\end{figure}
It has been proposed from experimental data below 1 GPa, that the
evolution of the Lorentz-Lorenz factor of various molecular systems
should follow an universal curve when plotted in reduced
units~\cite{Lallemand77}, after small statistical dipole-induced dipole
corrections. In our case, their relative effect on $LL$ is smaller than
10$^{-3}$ and can be neglected. In figure~\ref{fig:LL}, the change of
the Lorentz-Lorenz factors of He, Ne, H$_2$, H$_2$0 vs. density is
plotted in reduced units, i.e. as $LL/LL_{gas}$ (because $LL_{gas}$,
taken in~\cite{Achtermann93},~\cite{Lallemand77},~\cite{Schiebener90},
contains the isolated atom polarizability) vs. $N_A \rho \sigma^3$,
(where $\sigma$ is the Lennard-Jones parameter of the considered
compound). The following Lennard-Jones parameters have been used: He,
$\sigma=2.556$ \AA~\cite{Klein76}; Ne, $\sigma=2.83$ \AA~\cite{Vos91};
H$_2$, $\sigma=3.06$ \AA~\cite{Ross83}; H$_2$O, $\sigma=2.85$
\AA~\cite{Belonoshko91}. Calculations have been performed using the
following published EoS: He~\cite{LeToullec89},~\cite{Loubeyre93},
Ne~\cite{Vos91},~\cite{Lallemand77},
~\cite{Finger81},~\cite{Kortbeek88}, and
H$_2$\cite{Mills77},~\cite{Hemley90},~\cite{Pratesi97}. The scatter of
the points give an idea of the experimental errors. The large scatter
at low pressure observed for He and H$_2$, the most compressible
materials, is mainly caused by the uncertainty on density calculation;
moreover, uncertainty on $LL$ is larger when $n$ is close to one. At
higher compression, relative uncertainty decreases.

In Fig.~\ref{fig:LL}, a common trend is observed, namely a decrease of
the polarizability with density for the four systems. In fact, the
polarizability is related to the extent of the electronic cloud. In a
dense environment, the atoms and molecules themselves compress to lower
their interactions with their neighbors, and a decrease of the
polarizability is expected. Over the compression range covered here,
the scaled $LL$ factors of H$_2$ and of He, two isoelectronic species,
exhibit the same density evolution. However, in that pressure range,
the fundamental difference between the density effect on electronic
properties of these two media would be extracted from dispersion data.
In He, it has been shown that the main contribution of the change of
polarizability with density is taken into account by the blue shift of
the 1S-2P transition energy~\cite{LeToullec89}. In contrast, the change
of polarizability in hydrogen has been related to the red shift of the
excitonic level $X^{1}\Sigma_{g}^{+}$ to $B^{1}\Sigma_{u}^{+}$ that
should lead to the metal
hydrogen~\cite{VanStraaten88b},~\cite{Garcia92}. But different
evolutions are observed in Ne and H$_2$O; the proposed universal
behavior~\cite{Lallemand77} is nor valid nor theoretically justified
for a large perturbation by density.
In the past 20 years, numerous computations and measurements have been
performed for rare gas fluids
(see~\cite{Dacre81},~\cite{Achtermann93},~\cite{Hattig99} and
references therein), in order to reproduce, with success for rare-gas
 fluids, the two and three order refractivity virial coefficients.
 Yet, they have not been extended in the
dense fluid or the solid phases. We are not aware of any calculation on
the dielectric properties of water under pressure. On the other hand,
the calculation of the dielectric properties of hydrogen is a subject
of great current interest (see~\cite{Souza00} and references therein).
We present here an interesting database of the optical dielectric
properties of low Z compounds under pressure. We hope that it could be
helpful to validate {\it ab-initio} approaches.


\subsection{Gladstone-Dale relation }
\begin{figure}[tp]
\includegraphics[height=0.7\linewidth,clip=]{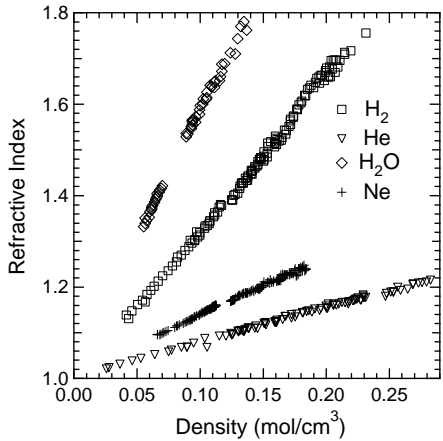}   
\caption{Refractive index vs. density of the studied compounds. The
corresponded fitted linear functions are on table~\ref{IndexVsRhoTable}
\label{fig:GD}}
\end{figure}

\begin{table}[htb]
\begin{tabular}{ccc}
\hline
Compound  & $a$ &$b$ (cm$^3$/mol)  \\
\hline
He &$1.005 \pm 0.001$ & $  0.749\pm  0.006 $   \\
Fluid Ne&  $1.000  \pm 0.001  $&  $ 1.413 \pm 0.009  $\\
Solid Ne&  $ 1.011 \pm 0.003  $&  $  1.25\pm  0.02 $\\
Fluid H$_2$& $   0.994\pm 0.003  $&  $ 3.26 \pm 0.03  $\\
Solid H$_2$&  $ 0.953 \pm 0.007  $&  $ 3.53 \pm  0.04 $\\
Fluid H$_2$O   & $  1.00\pm  0.01 $ & $  6.05\pm0.2   $\\
Ice VII & $ 1.081 \pm 0.01  $ & $  5.083\pm 0.13  $\\
\hline
\end{tabular}
 \caption{Density linear evolution of the refractive index obtained in
 this work ; the index expresses as:
 $n=a+b\rho$.\label{IndexVsRhoTable}}
\end{table}

The Gladstone-Dale relation expresses as~\cite{Setchell02}:
\begin{equation}
\frac{\mathrm{d} n }{ n-1} =\frac{\mathrm{d}\rho}{\rho}, \label{eq:GD}
\end{equation}
which leads to $n=1+a \rho$. It can be derived by differentiating Eq.
\ref{LorLor} and expanding it as a function of $u=n-1$, if the
polarizability is assumed to remain constant :
$$
\frac{\mathrm{d} u}{ u} \left( 1 - \frac{1}{6}u -\frac{5}{6} u^2
+...\right)=\frac{\mathrm{d}\rho}{\rho}.
$$
Thus, this relation corresponds to the zero-order truncated series
expansion of $\mathrm{d} u$ vs. $\mathrm{d} \rho$. For $n<1.8$, the
first order term remains smaller than 13 \% of the zero-order term.
Moreover, the volume dependency of $\alpha$ will tend to compensate the
effect of the first and second order terms.

The Gladstone-Dale relation has been widely used to analyze the
shock-wave VISAR data. In fact, a transparent window is often attached
to the rear face of the sample, together whith an optically reflecting
interface, to prevent the formation of a reflected rarefaction wave.
The high pressure refractive index of this window is needed to
determine the rear face velocity of the sample by VISAR
technique~\cite{Setchell02}. It can also been useful if the sample
itself is transparent, and if the shock front becomes reflective. This
is the case for pre-compressed samples~\cite{Lee02}. An advantage of
the Gladstone-Dale relation is that refractive index expresses as a
function of the density of the material alone (intrinsic temperature
effect is not taken into account), and very simply. In this peculiar
form, it allows to express straightforwardly the actual velocity of the
reflecting interface as a function of its apparent velocity, for a
planar single shock wave.

For this reason, we think that a presentation of our results in a form
of Glastone-Dale fit could be useful for the high pressure community.
They are presented in Fig.~\ref{fig:GD}. The data are close to straight
lines, even if a small curvature is obvious in all curves. The
Gladstone-Dale relation is often presented under the following form:
$n=a+b\rho$~\cite{Setchell02}. We used this function to fit our data.
The results are presented on Table~\ref{IndexVsRhoTable}.

\section{Applications}

\subsection{Equation of State}
\begin{figure}[tp]
\includegraphics[height=0.85\linewidth,clip=]{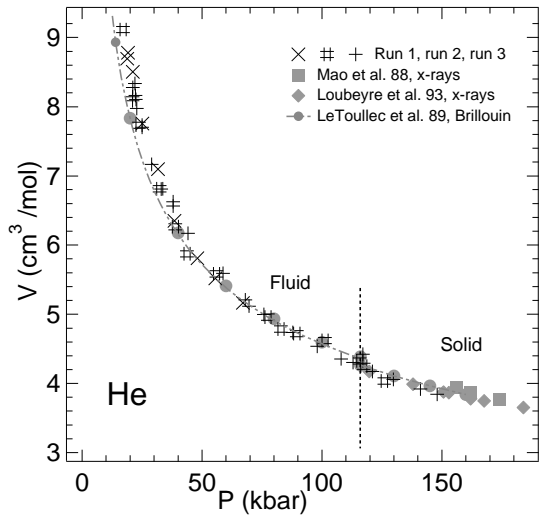}   
\caption{EoS of helium deduced from the sample thickness evolution upon
pressure increase for runs 1, 2 and 3. It compares very well with
previous determinations (\protect\cite{Mao88},~\protect\cite{LeToullec89},~\protect\cite{Loubeyre93}), within 1 \%,
except for the lowest pressures. \label{fig:EoSHe}}
\end{figure}
The EoS of helium in the fluid and solid phase has been measured up to
13 GPa, as presented in Fig.~\ref{fig:EoSHe}, during runs 1 to 3. The
volume of the sample was obtained by the multiplication of its
thickness by its surface. Sample thickness was varying between 37
$\mu$m and 23 $\mu$m (simple gaskets).
\begin{figure}[tp]
\includegraphics[height=0.65\linewidth,clip=]{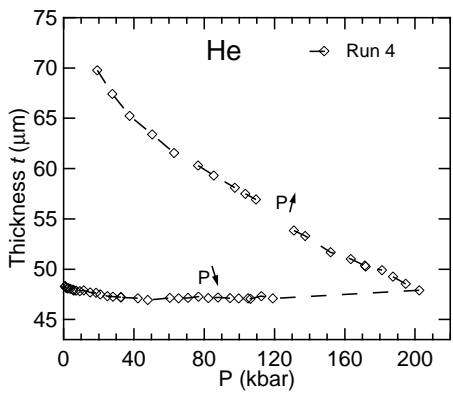}   
\caption{Evolution of the sample thickness upon pressure increase and
decrease for Run 4.\label{fig:ThickHe}}
\end{figure}
The evolution of the sample thickness during run 4 (double rhenium
gasket) is also plotted in Fig.~\ref{fig:ThickHe}. It is interesting to
note
 that the thickness of the sample decreased non-linearly upon pressure
increase but remained approximately constant upon pressure decrease:
from 47 $\mu$m to 48.5 $\mu$m. This can be used experimentally, when
the sample thickness knowledge is needed~\cite{Shen02}: one can perform
measurements upon pressure decrease.

Our EoS is compared with Brillouin spectroscopic
measurements~\cite{LeToullec89} in the fluid phase and x-ray data in
the solid phase~\cite{Mao88},~\cite{Loubeyre93}. At moderate pressures
($P < 3$ GPa), it seems that the compressibility of the sample is
overestimated; this effect can be attributed to either leakage of the
sample or geometrical effects. Between 3 and 14 GPa, agreement with
other data is excellent. No evident leakage has been observed in that
pressure range. This shows that this method for EoS measurements is
very interesting for liquids, where x-ray diffraction measurements are
still difficult to perform and where there is no radial pressure
gradient in the sample. In particular, these measurements could be
carried out for high temperature and high pressure liquids.

\subsection{Detection of phase changes}

The volume discontinuities at each phase changes can be deduced from
measurement. In fact, we observed no evidence the the $LL$ factor was
discontinuous at any phase change studied. Thus, we assumed that this
function depends on density alone. By differentiating Eq.~\ref{LorLor},
$\Delta \rho/\rho$ can then be deduced from the observed $\Delta n$.
The results are summarized table~\ref{TableDeltaV}. Our data compare
very well with those obtained by other techniques. Thus, careful
refractive index measurement can detect phase changes with volume
changes as small as 2 \%.





\begin{table}[htb]
\begin{tabular}{ccccc}
\hline
Phase change  & $P$ (kbar) & $\Delta n $  & $\Delta \rho/\rho$  & $\Delta \rho/\rho$  \\
  & &  $\times 10^3$&   (\%)& literature (\%) \\
  \hline
Fusion He & 116 & 3 $\pm$ 2  & 2 $\pm$ 1.3 & 3 \cite{LeToullec89}, \cite{Loubeyre93}\\
Fusion Ne & 47.7 & 6 $\pm$ 2  & 3.5 $\pm$ 1.2 & 3.8 \cite{Vos91}\\
Fusion H$_2$  &  53 &  8 $\pm$ 2&  3 $\pm$ 0.8& 5 \cite{Mills77}, \cite{Pratesi97}\\
Fusion H$_2$O   & 9.7 & 41 $\pm$  4&  9.7 $\pm$ 1 & 8.9 \cite{Bridgman42}\\
Ice VI $\rightarrow$ VII &  21.5 &37 $\pm$  4  & 8.0 $\pm$ 0.8 & 8.4   \cite{Bridgman42} \\
\hline
\end{tabular}
\caption{Density discontinuities at phase changes; the temperature of
measurements was between 297 and 299 K.\label{TableDeltaV} }
\end{table}

\section{Conclusion}

We have performed accurate optical refractive index measurements of
H$_2$O, He, H$_2$ and Ne up to 35 GPa and at ambient temperature in a
diamond anvil cell. These data should b useful to analyze some
spectroscopic measurements on these systems at high pressure. The
change of the refractive index has been expressed in terms of the
change of the polarizability of the molecular entities in their dense
surroundings. It is still challenging to compute such electronic
properties. The present set of data could be useful in validating the
various theoretical approaches. Our method also allows the direct
measurements of the Equations of State of transparent media: this has
been carried out for fluid and solid He, with less than 1 \% error.
This method should then be interesting to determine the equation of
state of dense fluids or dense amorphous media.

\acknowledgements{We thank M. Berhanu for his help during Neon
experiments.}


\end{document}